\documentclass[12pt]{article}

\usepackage{amsmath}
\usepackage{amssymb}
\usepackage{graphicx}
\usepackage{color}

\newcommand{\psib}{\bar{\psi}}

\newcommand{\nn}{\nonumber}

\newcommand{\vp}{\varphi}
\newcommand{\der}{\partial}
\newcommand{\vpb}{\bar{\varphi}}
\newcommand{\im}{\mbox{Im}}
\newcommand{\al}{\alpha}
\newcommand{\be}{\beta}
\newcommand{\jb}{\bar{j}}
\newcommand{\vpA}{\varphi^{\scriptscriptstyle{A}}}
\newcommand{\vpB}{\varphi^{\scriptscriptstyle{B}}}
\newcommand{\vpC}{\varphi^{\scriptscriptstyle{C}}}
\newcommand{\vpAb}{\bar{\varphi}^{\bar{\scriptscriptstyle{A}}}}
\newcommand{\vpBb}{\bar{\varphi}^{\bar{\scriptscriptstyle{B}}}}
\newcommand{\vpCb}{\bar{\varphi}^{\bar{\scriptscriptstyle{C}}}}
\newcommand{\sa}{\scriptscriptstyle{(a)}}

\newcommand{\ba}{\begin{eqnarray}}
\newcommand{\ea}{\end{eqnarray}}

\makeatletter 

\@addtoreset{equation}{section}
\makeatother

\begin{document}


\thispagestyle{empty}

\begin{flushright}
{\large KUNS 1944}
\end{flushright}
\vspace{20mm}
\begin{center} 
{\bf \Large  Construction of Supergravity Backgrounds\\ \vspace{8pt} with a Dilaton Field}
\vspace{20mm}

  {\bf Noriko Nakayama}
\footnote{\it  e-mail address: nakayama@phys.h.kyoto-u.ac.jp}\\
\vspace{8pt} 
{\it Graduate School of Human and Environmental Studies, Kyoto University, Kyoto 606-8501, Japan}\\ 
\vspace{10pt} 
and \\ 
\vspace{10pt} 
  {\bf Katsuyuki Sugiyama}
\footnote{\it  e-mail address: sugiyama@phys.h.kyoto-u.ac.jp} \\ 
\vspace{8pt}
{\it Graduate School of Science, Kyoto University, Kyoto 606-8501, Japan} 

\vspace{40pt}

{\bf Abstract}\\[5mm]
{\parbox{13cm}{\hspace{5mm}

A new class of 
non-compact K\"ahler backgrounds accompanied by a non-constant dilaton field
is constructed as a supergravity solution. 
It is interpreted as a complex line bundle over a base manifold comprising of
a combination of arbitrary coset spaces, and also includes
the case of Calabi-Yau manifolds.
The resulting backgrounds have $U(1)$ isometry. 
We consider $N=2$ supersymmetric $\sigma$-models on them,
and derive a non-K\"ahlerian solution by $U(1)$ duality transformation,
which preserves $N=2$ supersymmetry.

}}

\end{center}

\newpage
\section{Introduction} 
 
Non-critical string theories with a non-trivial dilaton field have been extensively studied 
in order to analyse dynamics of two-dimensional gravity and Liouville theory.
The Liouville theory is equivalent to the linear dilaton system,
where the configuration of the dilaton has rich structures 
in dynamics of string theories.

For the purpose of generalizing this dilaton system,
we construct supergravity backgrounds
interpreted as a complex line bundle over a base manifold comprising of
a combination of arbitrary coset spaces.
They have a non-constant dilaton field in general. 
Our study includes the discussion in Ref.\cite{Higashijima:2002px}, 
which solves the Einstein equation to get Ricci flat K\"ahler metrics of
a complex line bundle over coset spaces.

The backgrounds found here necessarily have $U(1)$ isometry.
By considering an $N=2$ supersymmetric $\sigma$-model on them, we can
perform $U(1)$ duality transformation.\cite{Kiritsis:1993pb}
The duality symmetry originating from such an isometry can be understood as
replacing a chiral superfield with a twisted chiral superfield by a Legendre transformation.
The duality symmetry also relates backgrounds which have geometrically 
different properties but correspond to the same conformal field theory.
Therefore, by this transformation, we can get another $\sigma$-model, which is equivalent to the original 
one as a conformal field theory.
The resulting dual background consists of a metric, a dilaton, and an anti-symmetric tensor field.
As a result it is no longer K\"ahlerian possessing non-trivial torsion. 

This paper is organized as follows:
In Section \ref{const_metric}, we explain a complex line bundle over coset spaces,
and then derive a non-compact K\"ahler background with a non-trivial dilaton as a solution of supergravity.
In Section \ref{example}, we give several explicit examples of such backgrounds.
In Section \ref{duality_tr}, we consider the $N=2$ supersymmetric $\sigma$-model 
on the background obtained in Section \ref{const_metric}. By performing a duality transformation on it,
we construct a non-K\"ahlerian background. 
Section \ref{conclusion} is devoted to summaries and conclusions.

\section{Complex Line Bundle over Coset Spaces}
\label{const_metric}

In this section we construct a new class of K\"ahler metrics with a non-trivial dilaton.
We consider a
canonical line bundle $L$ over an arbitrary K\"ahler coset space $M$,
and couple the metric $G_{MN}$ to the dilaton $\Phi$.
The base manifold $M$ is a direct product of $N$
K\"ahler coset spaces $G_a/H_a$ $(a=1,\cdots, N)$
\ba
M=\left(G_1/H_1\right)\times \left(G_2/H_2\right)\times 
\cdots \times
\left(G_N/H_N\right),
\ea
and we assume the K\"ahler potential $K$ of $L$ is  a function of the $G$-invariant $X$ defined
by\cite{Higashijima:2002px}
\ba
&&X=\log\left(
|\sigma|^2 e^{\sum^N_{a=1}h_a\Psi_a}\right)
=\log |\sigma|^2+\hat{\Psi}, \quad 
\hat{\Psi}=\sum^N_{a=1}h_a\Psi_a,
\label{def_x}
\ea
where $\Psi_a$ is a K\"ahler potential corresponding to $G_a/H_a$, $h_a$ is a real positive constant,
and $\sigma$ is a complex coordinate of a fiber of $L$.

The technology for finding the K\"ahler potential for an arbitrary coset space $G_a/H_a$ is developed  
in Refs.\cite{Itoh:jz,Itoh:ha},
using the supersymmetric nonlinear realization \cite{Bando:1983ab,Bando:1984cc,Bando:1984fn}.
Let $G_a^{\bf C}$ be a complexification of $G_a$, and
$\hat{H}_a$ be some complex group including $H_a^{\bf C}$ (a complexification of $H_a$)
such that Lie algebra $\hat{\cal H}_a$ of $\hat{H}_a$ is complex isotropy algebra.
There is a homeomorphism $G_a/H_a \cong G_a^{\bf C}/\hat{H}_a$.
Complex coordinates $\vp^i$ parametrize the coset space $G^{\bf C}_a/\hat{H}_a$, and
the representative of the coset is given by
\begin{eqnarray}
	\xi_a(\vp) = \exp (i \vp \cdot {\cal{X}}), \quad \{ {\cal{X}} \}:\mbox{generators of}\ \  
	{\cal G}_a^{\bf C} - \hat{\cal H}_a.
\end{eqnarray}
The K\"ahler potential of $G_a/H_a$ is represented as
\ba
 &&\Psi_a(\varphi ,\bar{\varphi})=\sum^{k_a}_{\alpha_a =1}v_{\alpha_a}
\log\det_{\eta_{\alpha_a}}\xi^\dagger_{a}(\vpb) \, \xi_a (\vp),
\ea
where $k_a$ is the dimension of torus in $H_a$, 
$v_{\alpha_a}$ is a real positive constant, 
and $\det_{\eta_{\alpha_a}}$  
is the determinant of the subspace projected by a projection matrix $\eta_{\alpha_a}$.
This yields the K\"ahler metric and the Ricci tensor
\ba
&&g^{\sa}_{i\bar{j}}=\der_i\der_{\bar{j}}\Psi_a,
\quad
R^{\sa}_{i\bar{j}}=-\der_i\der_{\bar{j}}\log\det g^{\sa}_{k\bar{\ell}},
\label{r1}
\ea
where $\der_i = \frac{\der}{\der \vp_i}$.
With a suitable choice of $v_{\alpha}$, any K\"ahler coset space is shown to become a
K\"ahler-Einstein manifold
\ba
&&R^{\sa}_{i\bar{j}}=h_a g^{\sa}_{i\bar{j}}.
\label{r2}
\ea
Note that 
$h_a$ in Eq.(\ref{def_x}) is the same as this coefficient.
Eqs.(\ref{r1}) and (\ref{r2}) give the expression of the determinant
with a holomorphic function ``$hol.$"
\ba 
\det g^{\sa}_{i\bar{j}}=e^{-h_a\Psi_a}\cdot |hol.|^2.
\ea

The total space $L$ is parametrized by the coordinates
$\sigma,\,\varphi^i_a \, (a=1,2,\cdots ,N)$,
where the index $i$ runs through the dimension of each coset space as $i=1,\cdots, \dim_{C} G_a/H_a$,
but we often omit ``$a$" if there is no confusion.
The metric $G_{MN}$ of $L$ is given in terms of the K\"ahler  potential $K$ as
$G_{\mu\nu}=G_{\bar{\mu}\bar{\nu}}=0$ and
\ba
&&G_{\mu\bar{\nu}}=\der_{\mu}\der_{\bar{\nu}}K
=\left(
\begin{array}{cc}
G_{\sigma\bar{\sigma}} & G_{\sigma\bar{j}}\\
G_{i\bar{\sigma}} & G_{i\bar{j}}
\end{array}
\right),
\ea
whose components are represented as
\ba
&&G_{\sigma\bar{\sigma}}=K''\frac{\der X}{\der\sigma}
\frac{\der X}{\der\bar{\sigma}},\quad
G_{i\bar{j}}=K''\frac{\der X}{\der\varphi^i}
\frac{\der X}{\der\bar{\varphi}^{\bar{j}}}
+K'\frac{\der^2K}{\der\varphi^i\der\bar{\varphi}^{\bar{j}}},\\
&&
G_{i\bar{\sigma}}=K''\frac{\der X}{\der\bar{\sigma}}
\frac{\der X}{\der \varphi^i},\quad
G_{\sigma\bar{j}}=K''\frac{\der X}{\der\sigma}
\frac{\der X}{\der\bar{\varphi}^{\bar{j}}},
\ea
where $K'=dK/dX$, and the Ricci-tensor is
\begin{eqnarray}
R_{\mu\nu}=R_{\bar{\mu} \bar{\nu}}=0, \quad	
R_{\mu\bar{\nu}}= -\der_{\mu}\der_{\bar{\nu}}
\log\det G_{\kappa\bar{\lambda}}. 
\label{general_ricci}
\end{eqnarray}
In order that this background provides a consistent string theory,
the two-dimensional $\sigma$-model has to be conformal invariant.
At one-loop level, conformal invariance requires 
the following equations of motion for the background fields $G_{MN}$ and $\Phi$:
\begin{eqnarray}
&&R_{\mu\bar{\nu}}=-2\der_{\mu}\der_{\bar{\nu}}\Phi,\label{beta1}\\
&&R_{\mu\nu}=-2\nabla_{\mu}\nabla_{\nu}\Phi.\label{beta2}
\end{eqnarray}

Let us solve these differential equations to obtain the metric
and the dilaton under the ansatz $K=K(X)$, $\Phi=\Phi(X)$.
Eqs. (\ref{general_ricci}) and (\ref{beta1}) imply 
\ba
&&
\det G_{\mu\bar{\nu}}
=e^{2\Phi}|hol.|^2.
\label{beta1''}
\ea
From the expression 
\begin{eqnarray}
&&\det G_{\mu\bar{\nu}}=
\det G_{\sigma\bar{\sigma}}\cdot\det (G_{i\bar{j}}-G_{i\bar{\sigma}}
\frac{1}{G_{\sigma\bar{\sigma}}}G_{\sigma\bar{j}})\nn\\
&&\hspace{37pt} =e^{-X}K''\cdot (K')^D\cdot |hol.|^2
=\frac{1}{D+1}e^{-X}\frac{d}{dX}(K')^{D+1}\cdot |hol.|^2,
\end{eqnarray}
Eq.(\ref{beta1''}) becomes
\begin{eqnarray}
&&2\Phi =-X+\log \left[
\frac{1}{D+1}\frac{d}{dX}(K')^{D+1}
\right],
\label{beta1'}
\end{eqnarray}
where we defined
\begin{eqnarray}
	D= \sum_{a=1}^{N} \dim_C G_a/H_a.
\end{eqnarray}
Also Eqs. (\ref{general_ricci}) and (\ref{beta2}) 
imply
\ba
\nabla_{\mu}\nabla^{\bar{\nu}}\Phi =
\der_{\mu}(G^{\bar{\nu}\lambda}\der_{\lambda}\Phi)=0.
\label{beta2''}
\ea
From the expressions
\begin{eqnarray}
&&G^{\bar{\nu}\mu}
=\left(
\begin{array}{cc}
G^{\bar{\sigma}\sigma} & 
G^{\bar{\sigma}i}\\
G^{\bar{j}\sigma} & G^{\bar{j}i}
\end{array}
\right)
=\left(
\begin{array}{cc}
\frac{|\sigma|^2}{K''}+\frac{|\sigma|^2}{K'}
\displaystyle{\sum_a} h_ag_{\sa}^{\bar{j}i}\der_{\bar{j}}\Psi_a\cdot \der_i\Psi_a &
-\frac{\bar{\sigma}}{K'}\displaystyle{\sum_a} \der_{\bar{j}}\Psi_a\cdot
      g^{\bar{j}i}_{\sa}\\ 
 -\frac{\sigma}{K'}\displaystyle{\sum_a} g^{\bar{j}i}_{\sa}\cdot \der_i\Psi_a
& \frac{1}{K'}\displaystyle{\sum_a} h^{-1}_ag^{\bar{j}i}_{\sa}
\end{array}
\right),\nn\\
\end{eqnarray}
Eq.(\ref{beta2''}) becomes 
\begin{eqnarray}
	\der_{\mu}(G^{\bar{\sigma}\lambda}\der_{\lambda}\Phi)
=\bar{\sigma}\der_{\mu}[(K'')^{-1}\der_X\Phi]=0,
\end{eqnarray}
so we get the equation
\begin{eqnarray}
	\der_X[(K'')^{-1}\der_X\Phi]=0.
\end{eqnarray}
With the definition of the variable $Y$
\begin{eqnarray}
	Y=K',
	\label{def_y}
\end{eqnarray}
this differential equation can be solved to yield the dilaton
\ba
2\Phi =-aY+b, \quad a,b=const., \quad a \geq 0.
\label{dilaton}
\ea
In the case of $a=0$, 
the background with a constant dilaton becomes a Calabi-Yau manifold 
due to Ricci flatness and our discussion 
coincides with Ref.\cite{Higashijima:2002px}.
In the case of $a \not= 0$, 
the background corresponds to the linear dilaton system with respect to $Y$.

For $a \not= 0$ we substitute (\ref{dilaton}) into (\ref{beta1'}) and acquire
\ba
e^{X+b}=\int^Y_0dY\,Y^De^{aY}+C,
\label{solution}
\ea
with an integration constant $C$.
By using the formula 
\ba
&&\int^y_0dx\,x^ne^x=(-1)^nn!
\left[
-1+e^y\sum^n_{m=0}\frac{(-y)^m}{m!}
\right]\,,
\ea
Eq.(\ref{solution}) can be rewritten into
\ba
&&Be^X=
-A+e^{aY}\sum^D_{m=0}\frac{(-aY)^m}{m!}, \label{solution_beta}\\ 
&& A=1- C\cdot (-1)^D a^{D+1} \cdot \frac{1}{D!},\quad
B=e^b\cdot (-1)^D a^{D+1}\cdot\frac{1}{D!}.
\ea
Thus we obtain the metric of the total space $L$.
Using Eqs.(\ref{def_y}) and (\ref{solution}), and setting $\sigma = e^{\frac{r}{2}+i\theta}$
the metric takes the form
\ba
&&ds^2=G_{MN}dx^Mdx^N=2G_{\mu\bar{\nu}}dx^{\mu}dx^{\bar{\nu}}\nn\\
&& \hspace{20pt} = \frac{a}{2f(aY)}dY^2+\frac{2f(aY)}{a}
\left[\mbox{Im}(d\varphi^i\der_i\hat{\Psi})
+ d \theta \right]^2
+2Y\sum_{a=1}^{N}h_ads_{a}^2,
\label{metric}
\ea
with 
\begin{eqnarray}
&&ds_{a}^2=g_{i\bar{j}}^{\sa}d\varphi^id\bar{\varphi}^{\bar{j}},\\
&&f(aY)=aY'=(-1)^D D!(aY)^{-D}
\left[
-Ae^{-aY}+\sum^D_{m=0}\frac{(-aY)^m}{m!}
\right],\\
&& d \theta =\im (d \log \sigma).
\end{eqnarray}	
The scalar curvature of this background metric can be calculated as 
\ba
&&R=2a(1-aY')=2a(1-f(aY)).
\ea
We can evaluate asymptotic behavior of the function $f(aY)$
\begin{eqnarray*}
	&&f(aY) \approx 1-\frac{D}{aY}
	-A\frac{(-1)^{D}D!}{(aY)^D} e^{-aY}+ {\cal O}\left( \frac{1}{(aY)^2} \right) \quad (aY \gg 1),\\
	&&f(aY) \approx (-1)^D D!\, \frac{1-A}{(aY)^D}e^{-aY} +
	    \frac{aY}{D+1} + {\cal O}\left( (aY)^2 \right) \quad (aY \approx 0).
\end{eqnarray*}
Therefore in the limit of $Y \rightarrow \infty$, $R$ approaches $0$.
At $Y=0$, the space is regular for $C=0$ ($A=1$),
but it has a singularity for $C\not=0$ ($A\not=1$).

For $a=0$, 
Eq.(\ref{solution}) leads to
\begin{eqnarray}
	\frac{f(aY)}{a}= Y'= \frac{1}{D+1}Y+ C\cdot Y^{-D},
\label{calabi}
\end{eqnarray}
and by plugging this into Eq.(\ref{metric}) we can get the expression of the metric.

Moreover the central charge deficit $\delta c$ provided by this background is
determined by the $\beta$-function of the dilaton field as
\begin{eqnarray}
	\delta c =c-\frac32\cdot 2(D+1) =  3 \left[ 2 (\nabla \Phi )^2 - \nabla^2 \Phi \right].
\end{eqnarray}
By inserting the dilaton (\ref{dilaton}) this becomes
\begin{eqnarray}	
	\delta c=\left[ a f(aY) + a f'(aY) + \frac{D}{Y}f(aY)\right] = 3 a,
	\label{cc}
\end{eqnarray}
and we obtain the central charge
\begin{eqnarray}
	c=3(D+1+a).
\end{eqnarray}

\section{Examples}
\label{example}

As discussed in the previous section,
once we choose the base coset space $M$,
we can construct the metrics of the complex line bundle over the base manifold;
calculating $\im(d \vp^i \der_i \hat{\Psi})$ and $\der_i\der_{\jb}\hat{\Psi}$  
yields background metrics (\ref{metric}).
In this section, we give some concrete examples by using K\"ahler potentials for various coset spaces.

In the following subsections,
we consider only the case that the base manifold consists of a single coset space,
since the discussion on the case containing more than one K\"ahler coset spaces is straightforward.

\subsection{Hermitian Symmetric Spaces}
To begin with, 
let us construct metrics for the complex line bundles over hermitian symmetric spaces. 
For these case, 
several examples of the K\"ahler potentials for the base coset spaces 
are explicitly derived in Refs.\cite{Itoh:jz,Higashijima:1999ki},
and
the $G$-invariant $X$ 
is expressed by \cite{Higashijima:2002px}
\begin{eqnarray}
X=\log |\sigma|^2 + \hat{\Psi},\quad
\hat{\Psi}= h \Psi,\quad
\Psi= v\log\det_{\eta}\xi^\dagger \xi \equiv v \log \Xi, \quad 
h= \frac{1}{2 v}\, 2\, \tilde{h}(G),
\end{eqnarray}
where $\tilde{h}(G)$ is the dual Coxeter number of $G$, and we normalized the generators 
of the fundamental representation as
\begin{eqnarray}
	\mbox{Tr}(T^A T^B)= \delta^{AB}.
\end{eqnarray}  
We summarize their results in Table.\ref{hss}.
From these results we calculate the metrics for the complex line bundles over the hermitian symmetric spaces.  
\begin{table}[h]
\caption{Summary of hermitian symmetric spaces $G/H$.}
\begin{center}
\renewcommand{\arraystretch}{1.3}
\begin{tabular}{l l l l l}
	\hline
	$G/H$ &$\tilde{h}(G)$ &$\Xi$ & $\dim_{C} G/H$ \\
	\hline
	$CP^{N-1}$ & $N$ &$1+|\vp^i|^2$  & $N-1$ \\
	& & $(i=1,2,\cdots,N-1)$ & \\
	$Q^{N-2}$ & $N-2$ &$1+|\vp^i|^2+\frac{1}{4}(\vp^i)^2(\vpb^{\bar{j}})^2$ & $N-2$ \\
	& & $(i=1,2,\cdots,N-2)$ & \\
	$G_{N,M}$ & $N$ &$\det ({\bf 1}_M+\vp^{\dagger}\vp)$  & $M(N-M)$ \\
	& & $(\vp_{Aa}:A=1,2,\cdots,N-M,\ a=1,2,\cdots,M)$  & \\
	$Sp(N)/U(N)$ & $N+1$ &$\det ({\bf 1}_N+\vp^{\dagger}\vp)$  & $\frac12 N(N+1)$ \\
	& & $(\vp_{ab}: a,b=1,\cdots, N,\ \vp_{ab}=\vp_{ba},\  a\leq b)$ & \\
	$SO(2N)/U(N)$ & $2N-2$ &$\det ({\bf 1}_N+\vp^{\dagger}\vp)$  & $\frac12 N(N-1)$ \\
	& & $(\vp_{ab}: a,b=1,\cdots, N,\ \vp_{ab}=-\vp_{ba},\ a<b)$ & \\
	$E_6/[SO(10) \times U(1)]$ & $12$ &$ 1+|\vp_{\al}|^2+\frac{1}{8}|\vp_\al (C\sigma^{\dagger}_A)^{\al\be} 
	\vp_\be|^2$  & $16$ \\
	& &  $(\alpha = 1,2,\cdots,16,\, A=1,2,\cdots,10)$ & \\
	$E_7/[E_6 \times U(1)]$ & $18$ &$1+|\vp^i|^2+\frac{1}{4}|\Gamma_{ijk}\vp^j \vp^k|^2
    +\frac{1}{36}|\Gamma_{ijk}\vp^i \vp^j \vp^k |^2$  & $27$ \\
	& & $(i=1,2,\cdots,27)$ & \\
	\hline
\end{tabular}
\end{center}
\label{hss}
\end{table}

\subsubsection{Projective Space: $CP^{N-1}= SU(N)/[SU(N-1) \times U(1)]$}
By using $\Xi$ and $\tilde{h}(G)$ in Table.\ref{hss}, we can calculate
\begin{eqnarray}
	&&\der_i \der_{\jb} \hat{\Psi}= h \der_i \der_{\jb} \Psi 
	=h g_{i\bar{j}}= N\Biggr[ \frac{1}{\Xi}\delta_{i j}-\frac{1}{\Xi^2}\vp^j \vpb^{\bar{i}}\Biggl],\\
	&&\im (d \vp^i \der_i \hat{\Psi}) =
	 N \cdot \frac{i}{2} \frac{1}{\Xi}
	\left( d \vpb^{\bar{i}} \vp^i -  d \vp^{i} \vpb^{\bar{i}}  \right)\equiv -N A,
\end{eqnarray}
and we get the metric
\begin{eqnarray}
	ds^2= \frac{1}{2Y'}dY^2 + 2 Y'\left[ d \theta  -N A\right]^2 +2 NY ds_{FS}^2,
\end{eqnarray}
where
$A$ is a connection 1-form on $CP^{N-1}$ and $ds_{FS}^2$ is the Fubini-Study metric.
After rescaling $Y\rightarrow Y/N$, we obtain 
\begin{eqnarray}
	ds^2= \frac{1}{2NY'}dY^2 + \frac{2 Y'}{N}\left[ d \theta  -N A\right]^2 +2 Y ds_{FS}^2,
	\quad \Phi=- \frac{a}{2N}Y + const.
\end{eqnarray}
This corresponds to a solution of $\beta^G_{MN}=0$ under $U(N)$ isometry
found in Refs.\cite{Kiritsis:1993pb,Hori:2002cd,Nakayama:2004vy}
as a generalization of the two-dimensional black hole background \cite{Witten:1991yr,Mandal:1991ua}.
\\

For the other cases in Table \ref{hss},
we obtain the metrics by calculating $\der_i \der_{\jb}\hat{\Psi}$ and $\im (d\vp^i\der_i \hat{\Psi})$.
The results are collected in the following:

\subsubsection{Quadratic Space: $Q^{N-2}= SO(N)/[SO(N-2)\times U(1)]$}
\begin{eqnarray}
	&&\der_i \der_{\jb} \hat{\Psi}=(N-2) \Biggr[ \frac{1}{\Xi}( \delta_{ij} + \vp^i \vpb^{\bar{j}} )
	-\frac{1}{\Xi^2}(\vpb^{\bar{i}}+\frac12 \vp^i (\vpb^{\bar{k}})^2)
	(\vp^j+\frac12 (\vp^l)^2 \vpb^{\bar{j}})\Biggl],\\
	&&\im (d \vp^i \der_i \hat{\Psi}) =
	(N-2) \cdot \frac{i}{2} \frac{1}{\Xi}
	\Biggl[ d \vpb^{\bar{i}}(\vp^i+\frac12(\vp^k)^2 \vpb^{\bar{i}}) - d \vp^i(\vpb^{\bar{i}}+\frac12 \vp^i
	 (\vpb^{\bar{k}})^2) \Biggr],
\end{eqnarray}

\subsubsection{Grassmanian: $G_{N,M} = SU(N)/[SU(N-M) \times U(M)]$}

\begin{eqnarray}
   && \frac{\der^2 \hat{\Psi}} {\der \vp_{Aa} \der\vp^\ast_{Bb}} = N ({\bf 1}_M + \vp^\dagger \vp)^{-1}_{ab}
 \Biggl[ {\bf 1}_{(N-M)} -\vp ({\bf 1}_M+\vp^\dagger \vp)^{-1}\vp^\dagger \Biggr]_{BA}, \\
&&\im \left(  d \vp_{Aa}  \frac{\der}{\der \vp_{Aa}} \hat{\Psi}\right)=
N \cdot\frac{i}{2}({\bf 1}_M + \vp \vp^\dagger)^{-1}_{ab} (d \vp^\dagger \vp - \vp^\dagger d \vp)_{ba}.
\end{eqnarray}

\subsubsection{$Sp(N)/U(N)$ and $SO(2N)/U(N)$}

Setting $\epsilon= -1$ for $Sp(N)/U(N)$, and $\epsilon =+1$ for $SO(2N)/U(N)$, 
\begin{eqnarray}
	&& \frac{\der^2 \hat{\Psi}}{\der \vp_{ab} \der \vp^\ast_{cd}} = \tilde{h}(G)\ 
	\left(1 -\frac12 \delta_{ab}\right)\left(1- \frac12 \delta_{cd}\right)
	\Biggl[ 
	({\bf 1}_N +\vp^\dagger \vp)_{bd}^{-1} \{ 
	{\bf 1}_N -\vp ({\bf 1}_N+ \vp^\dagger \vp)^{-1}\vp^\dagger \}_{ca}\nn\\
	&&\quad \quad -\epsilon
	({\bf 1}_N +\vp^\dagger \vp)_{bc}^{-1} \{
	{\bf 1}_N -\vp ({\bf 1}_N+ \vp^\dagger \vp)^{-1}\vp^\dagger \}_{da}
+(a \leftrightarrow b,\ c \leftrightarrow d)
	\Biggr],\\
	&& \im \left( d \vp_{ab} \frac{\der \hat{\Psi} }{\der \vp_{ab}}\right)=
	\tilde{h}(G)\cdot \frac{i}{2} \left(1-\frac12 \delta_{ab}\right) \Biggl[ d \vp^\dagger_{ba}
	\{ (\vp ({\bf 1}_N +\vp^\dagger \vp)^{-1})_{ab}-\epsilon (\vp ({\bf 1}_N +\vp^\dagger \vp)^{-1})_{ba} \}\nn\\
	&&\quad \quad- d\vp_{ab}
	\{ (({\bf 1}_N +\vp^\dagger \vp)^{-1} \vp^\dagger)_{ba} - \epsilon (({\bf 1}_N +\vp^\dagger \vp)^{-1}\vp^\dagger)_{ab} \}
	\Biggr].
\end{eqnarray}

\subsubsection{Exceptional Group: $E_6/[SO(10) \times U(1)]$}

With 
$\vp_{\alpha}$'s  chiral superfields belonging to an $SO(10)$ Weyl spinor representation,
$\sigma_A$'s $SO(10)$ $\gamma$-matrices in the Weyl spinor basis,
and $C$ a charge conjugation matrix \cite{Higashijima:2001fp}, 

\begin{eqnarray}
&& \der_{\alpha} \der_{\bar{\beta}} \hat{\Psi} = 12 \Biggr[ \frac{1}{\Xi}\{   
\delta_{\al \be} + \frac12 (\sigma^A C^\dagger \vpb)^{\be} (C {\sigma_A}^\dagger \vp)^{\al} \}\nn\\
&&\quad\quad-\frac{1}{\Xi^2}\{   
\vpb_{\bar{\al}}+ \frac14 (C \sigma^\dagger_A \vp)^{\al} (\vpb \sigma^A C^\dagger \vpb)
\}\{  
\vp_{\be} + \frac14 (\sigma^A C^\dagger \vpb)^\be (\vp C \sigma^\dagger_A \vp)
\} \Biggr],\\
&&\im (d \vp^{\alpha} \der_{\alpha} \hat{\Psi})= 12 \cdot
\frac{i}{2} \frac{1}{\Xi}
\Biggl[ 
d \vpb_{\bar{\be}} \{ \vp_{\be} + \frac14 (\vp C {\sigma_A}^{\dagger} \vp)(\sigma^A C^{\dagger} \vpb)^{\be}\} \nn\\
&&\quad \quad-d \vp_\al \{ \vpb_{\bar{\al}} + \frac14 (\vpb \sigma^A C^\dagger \vpb)(C \sigma^\dagger_A \vp)^\al\}
\Biggr].
\end{eqnarray}

\subsubsection{Exceptional Group: $E_7/[E_6 \times U(1)]$}

With $\vp_i$'s chiral superfields belonging to the fundamental representation {\bf 27} of $E_6$,
$\Gamma_{ijk}$ the rank $3$ anti-symmetric tensor, and
$\Gamma^{ijk}$ its complex conjugate \cite{Higashijima:2001fp},

\begin{eqnarray}
&&\der_i \der_{\jb}\hat{\Psi}=
18 \Biggl[ \frac{1}{\Xi} \{ 
\delta_{ij} + (\Gamma_{ikl} \vp^l)(\Gamma^{jkm} \vpb_{\bar{m}})
+ \frac14 (\Gamma_{ikl} \vp^k \vp^l)(\Gamma^{jmn} \vpb_{\bar{m}} \vpb_{\bar{n}})
\} \nn\\
&&\quad\quad -\frac{1}{\Xi^2}\{ 
\vpb_{\bar{i}} + \frac12(\Gamma_{ikl} \vp^l)(\Gamma^{kmn} \vpb_{\bar{m}} \vpb_{\bar{n}}) 
+ \frac{1}{12}(\Gamma_{ikl} \vp^k \vp^l)(\Gamma^{mnp} \vpb_{\bar{m}} \vpb_{\bar{n}} \vpb_{\bar{p}})
\} \nn\\
&&\quad\quad \times \{ 
\vp^{j} + \frac12(\Gamma_{klm} \vp^l\vp^m)(\Gamma^{jkn} \vpb_{\bar{n}}) 
+ \frac{1}{12}(\Gamma_{klm} \vp^k \vp^l \vp^m)(\Gamma^{jnp} \vpb_{\bar{n}} \vpb_{\bar{p}})
\}\Biggr],\\
&&\im (d \vp^i \der_i \hat{\Psi}) = 18 \cdot
\frac{i}{2} \frac{1}{\Xi} \Biggl[ 
d \vpb_{\bar{i}} \{ \vp^{i} + \frac12 (\Gamma^{ilm} \vpb_{\bar{l}})(\Gamma_{mnp} \vp^n \vp^p)
+\frac{1}{12} (\Gamma^{ilm}\vpb_{\bar{l}} \vpb_{\bar{m}}) (\Gamma_{jnp} \vp^j \vp^n \vp^p)\}\nn\\
&&\quad \quad-
d\vp^i \{\vpb_{\bar{i}} + \frac12 (\Gamma_{ilm} \vp^l) (\Gamma^{mnp} \vpb_{\bar{n}} \vpb_{\bar{p}}) 
+\frac{1}{12} (\Gamma_{ilm}\vp^l\vp^m) (\Gamma^{jnp} \vpb_{\bar{j}} \vpb_{\bar{n}} \vpb_{\bar{p}}) \}
\Biggr].
\end{eqnarray}

\subsection{Non-Symmetric Spaces}

Next we consider $SU(l+m+n)/S[U(l)\times U(m) \times U(n)]$ 
derived in Ref.\cite{Itoh:jz},
as a example of non-symmetric spaces. 
It is known that there exists two kinds of complex structures on this coset 
space\cite{Buchmuller:1985cj,Buchmuller:1985rc}.
For simplicity, we explicitly construct the metric for the case of $l=m=n=1$.
In this case the two structures lead to the same model and the K\"ahler potential is expressed
with the complex coordinates $\vpA,\vpB,\vpC$ 
as\cite{Higashijima:2002px}\
\begin{eqnarray}
	&&\hat{\Psi}=h \Psi,\quad \Psi = v_1 \log \Xi_1 + v_2 \log \Xi_2,\quad h=\frac{2}{v_1}=\frac{2}{v_2},\\
      &&\Xi_1= 1 + |\vpC|^2 + |\vpB + \frac12 \vpA \vpC|^2,\quad
	\Xi_2 = 1 +|\vpA|^2 +|\vpB - \frac12 \vpA \vpC|^2. 
\end{eqnarray}
We can compute the metric in the same way as above
\begin{eqnarray}
	&&\der_i \der_{\jb}\hat{\Psi} = 
	2\Biggl[ 
	\frac{1}{\Xi_1{}^2}F(\vpA,\vpB,\vpC,\vpAb,\vpBb,\vpCb)
      +\frac{1}{\Xi_2{}^2}F(\vpC,-\vpB,\vpA,\vpCb,-\vpBb,\vpAb)
\Biggr],\\
&&\im (d\vp^i \der_i \hat{\Psi}) =
2 \cdot \frac{i}{2} \Biggl[ 
\frac{1}{\Xi_1}G(\vpA,\vpB,\vpC,\vpAb,\vpBb,\vpCb)+ \frac{1}{\Xi_2}G(\vpC,-\vpB,\vpA,\vpCb,-\vpBb,\vpAb)
\Biggr],\nn\\	
\end{eqnarray}
where $i=A, B, C$ and
\begin{eqnarray}
	&&F(\vpA,\vpB,\vpC,\vpAb,\vpBb,\vpCb)\nn\\
&&\   =\left(
\begin{array}{c}
d \vpAb \\
d \vpBb \\
d \vpCb
\end{array}
\right)^T
	\left(
	\begin{array}{ccc}
	 \frac14 |\vpC|^2 (1+|\vpC|^2) & \frac12 \vpCb (1+|\vpC|^2)  &
	 \frac14 \vpA\vpCb-\frac12 \vpB (\vpCb)^2  \\
	 \frac12 \vpC (1+ |\vpC|^2) & 1+|\vpC|^2 & \frac12 \vpA-\vpB \vpCb \\
	 \frac14 \vpAb \vpC -\frac12 \vpBb (\vpC)^2& \frac12\vpAb- \vpBb \vpC & 1+\frac14|\vpA|^2+|\vpB|^2
	\end{array}
	\right)
\left(
\begin{array}{c}
d \vpA \\
d \vpB \\
d \vpC
\end{array}
\right),\nn\\
&& \\
&&G(\vpA,\vpB,\vpC,\vpAb,\vpBb,\vpCb)\nn\\
&&\   =d\vpCb \vpC + d( \vpBb +\frac12 \vpAb \vpCb ) ( \vpB + \frac12 \vpA \vpC ) 
- d\vpC \vpCb - d(\vpB +\frac12 \vpA \vpC ) (\vpBb + \frac12 \vpAb \vpCb ).\nn\\
\end{eqnarray}
If $l,m,n$ take other values,
we can also obtain an explicit formula for the metric by the similar procedure.

\section{$U(1)$  Duality Transformation}
\label{duality_tr}

The K\"ahlerian backgrounds we have found are 
solutions of the equations of motion.
Since they have a $U(1)$ isometry, i.e. their metric $G_{MN}$ does not depend on $\theta$ explicitly,
we can construct their dual spaces by replacing one chiral superfield with 
a twisted chiral superfield\cite{Kiritsis:1993pb}.
The dual space in general contains non-trivial torsion 
and the resulting metric is no longer K\"ahlerian.
In this section, we construct the dual of the solution with a non-trivial dilaton 
obtained in Section \ref{const_metric}.

In our case the $N=2$ superspace action is determined by the following K\"ahler potential
\begin{eqnarray}
K=K(Z+\bar{Z}, \Phi_i, \bar{\Phi}_{\jb}),	
\end{eqnarray}
where $Z$ and $\Phi_i$ are chiral superfields whose lowest components are $z=\frac{r}{2} +i \theta
=\log \sigma$ 
and $\vp_i$ respectively.
In order to get the dual potential $\tilde{K}$, we perform the Legendre transformation as follows:
\begin{eqnarray}
	&&\tilde{K}(\Psi+\bar{\Psi}, \Phi_i, \bar{\Phi}_{\jb})=
	K(Z +\bar{Z},  \Phi_i, \bar{\Phi}_{\jb}) - (Z + \bar{Z}) (\Psi + \bar{\Psi}),\\
	&&\frac{\der K}{\der R}=\Psi +\bar{\Psi},\quad R =Z +\bar{Z},\label{legendre}
\end{eqnarray}
with a twisted chiral superfield $\Psi$ containing its lowest component $\psi$.
Now the independent variables are substituted
with $\psi, \psib, \vp_i, \vpb_{\jb}$.
The metric and anti-symmetric tensor are obtained by writing down the bosonic part of the superspace 
action \cite{Gates:1984nk},
\begin{eqnarray}
		\tilde{G}_{\mu \bar{\nu}}=
		\left(
		\begin{array}{cccc}
		 & -\tilde{K}_{\psi \bar{\psi}} &   & \\
		 -\tilde{K}_{\psi \bar{\psi}}&     &   \\
		 & & & \tilde{K}_{i \bar{j}}\\ 
		 & &\tilde{K}_{i \bar{j}}   &
		\end{array}
		\right), \quad
		B_{\mu \bar{\nu}}=
		\left(
		\begin{array}{cccc}
		 & & & \tilde{K}_{\psi \jb} \\
		 & &\tilde{K}_{i \bar{\psi}}  & \\
		 & -\tilde{K}_{i \bar{\psi}} & & \\ 
		-\tilde{K}_ {\psi \jb} & & &
		\end{array}
		\right),
\end{eqnarray} 
whose components are given by
\begin{eqnarray}
	&&\tilde{K}_{\psi \psib} = \frac{\der^2 \tilde{K}}{ \der \psi \der \bar{\psi} } 
	= -\frac{\der X}{\der \psi} = -\frac{a}{f(aY)}, \quad
      \tilde{K}_{i \jb}=\frac{\der^2 \tilde{K}}{\der \varphi^i \der \bar{\varphi}^{\bar{j}}} 
      = (\psi +\bar{\psi})\cdot \der_i \der_{\bar{j}} \hat{\Psi},\\
	&&\tilde{K}_{\psi \jb}=\frac{\der^2 \tilde{K}}{\der \psi \der \bar{\varphi}^{\bar{j}}}
	= \der_{\bar{j}} \hat{\Psi},\quad
	\tilde{K}_{i \psib}=\frac{\der^2 \tilde{K}}{\der \bar{\psi}\der \varphi^{i}}
	= \der_{i} \hat{\Psi}.
\end{eqnarray}
In the above calculation,
we used the relation derived from (\ref{def_y}) and (\ref{legendre})
\begin{eqnarray}
	Y = \psi +\bar{\psi},
\end{eqnarray}
and the expression of the derivative obtained from (\ref{solution_beta})
\begin{eqnarray}
	\frac{\der X}{\der \psi} = \frac{a}{f(aY)}.
\end{eqnarray}
Therefore we get the dual metric
\begin{eqnarray}
	d \tilde{s}^2 = \frac{2a}{f(aY)} d\psi d \bar{\psi} + 2 Y\,
	\der_i \der_{\jb} \hat{\Psi}\, d \vp^i d \vpb^{\bar{j}}.
\label{dual_metric}
\end{eqnarray}

The dual dilaton is discussed in Ref.\cite{Buscher:1987qj} to yield 
\begin{eqnarray}
	\tilde{\Phi} = \Phi - \frac12 \log \left( 2 K_{rr} \right),
\end{eqnarray}
where $r$ is the lowest component of $R=Z+\bar{Z}$,
and this turns out to be  
\begin{eqnarray}
\tilde{\Phi} 
= -\frac12 \log \left[ e^{a Y} \frac{f(a Y)}{a} \right] + const.
\label{dual_dilaton}
\end{eqnarray}
The dual dilaton $\tilde{\Phi}$ depends on $Y$ nonlinearly and gives nontrivial backgrounds.
We can calculate the field strength 
$H_{LMN}=\der_L B_{M N} + \der_M B_{NL} + \der_N B_{LM}$ and 
its non-zero components are represented by $\hat{\Psi}$
\begin{eqnarray}
H_{\psi i \jb}= -\der_i \der_{\jb}\hat{\Psi},\quad
H_{\psib i \jb}= \der_i \der_{\jb}\hat{\Psi}.	
\end{eqnarray}
These are collected into the $3$-form $H$ 
\begin{eqnarray}
	&&H= \frac16H_{LMN}\, dx^L \wedge dx^M \wedge dx^N 
	 = -2\, d(\im \psi)\wedge \sum_{a=1}^{N}h_a J_a,
\end{eqnarray}
where $J_a$ is the K\"ahler form of the coset $G_a/H_a$:
\begin{eqnarray}
	J_a = i g_{i \jb}^{\sa} d \vp^i \wedge d \vpb^{\jb}.
\end{eqnarray}
One can easily check $\tilde{G}_{MN}$, $\tilde{\Phi}$, and $H_{LMN}$
derived here satisfy the background field equations $\beta_{MN}^{\tilde{G}}=0$ and $\beta_{MN}^{B}=0$.
The central charge in this dual model is evaluated from $\beta^{\tilde{\Phi}}$ as
\begin{eqnarray}
	&&\delta \tilde{c}=\tilde{c}-\frac32\cdot 2(D+1) 
	=3 \left[ 2 (\nabla \tilde{\Phi})^2 - \nabla^2 \tilde{\Phi} 
	-\frac{1}{12}H_{LMN}H^{LMN}\right] \nn\\
      &&\hspace{13pt}=3\left[ a f(aY) + af'(aY) +\frac{D}{Y}f(aY)\right]=3a.
\end{eqnarray}
This result is equal to (\ref{cc}) completely, and consistent with the fact that 
the duality transformation preserves the conformal invariance of the theory.

The scalar curvature has a complicated form as
\begin{eqnarray}
	\tilde{R}&=& 2 \left[ 
      a\left( f''(aY) -\frac{f'(aY)^2}{f(aY)} \right) +\frac{3D}{2a}\frac{f(aY)}{Y^2}
      -\frac{D^2}{a}\frac{f(aY)}{Y^2}+\frac{D}{Y}
      \right]\nn \\
	&=&2 \left[ 
	a\left(1-\frac{1}{f(aY)} \right) + \frac{2 D}{Y} +\frac{5D}{2a}\frac{f(aY)}{Y^2}
	-\frac{D^2}{a}\frac{f(aY)}{Y^2} \right].
	\label{dual_r} 
\end{eqnarray}

Let us discuss the dependence on $Y$ of the dilaton $\tilde{\Phi}$ and the curvature $\tilde{R}$.
The $Y$ is limited to the region where the string coupling constant $g=\exp \tilde{\Phi}>0$.
For $C>0$ ($A>1$ for odd $D$,\ $A<1$ for even $D$) and $C=0$ ($A=1$),
$f(aY)>0$ in $Y>0$, and so $Y$'s range is $0<Y<\infty$.
For $C<0$ ($A<1$ for odd $D$,\ $A>1$ for even $D$ ),
there exists a unique $Y_0>0$ satisfying $f(aY_0) =0$, and $f(aY)$ monotonically increases in $Y>Y_0$.
Hence we consider $Y_0<Y<\infty$.
In Table.\ref{dependence}, we explain how $f(aY)$, $\tilde{\Phi}$, and $\tilde{R}$ vary depending on $Y$.
The space has one singular point and becomes flat as $Y \rightarrow \infty$ in each case. 
\begin{table}[h]
\caption{Dependence on $Y$ of $\tilde{\Phi}$ and $\tilde{R}$.  
  The ``$\ast$" means diverging to $+ \infty$ or $- \infty$ i.e. curvature singularity.}
\begin{center}
\renewcommand{\arraystretch}{1.3}
\begin{tabular}{c||c|c|c}
	\multicolumn{4}{l}{$C>0$}\\
	\hline
	$Y$     &$0$        & $\cdots$ & $\infty$  \\ \hline 
	$f(aY)$ & $+\infty$ & $\cdots$ & $1$ \\  \hline
	$\tilde{\Phi}$ & $-\infty$ &  & $-\infty$  \\
      $\tilde{R}$ &  $\ast$  &   & $0$ \\ \hline
\end{tabular}
\quad
\begin{tabular}{c||c|c|c}
\multicolumn{4}{l}{$C=0$}\\
	\hline
	$Y$     &$0$        & $\cdots$ & $\infty$  \\ \hline 
	$f(aY)$ & $0$ & $\cdots$ & 1\\  \hline 
	$\tilde{\Phi}$ & $+\infty$ &   & $-\infty$ \\ 
      $\tilde{R}$ &  $\ast$  &   & $0$ \\ \hline
\end{tabular}
\quad
\begin{tabular}{c||c|c|c}
\multicolumn{4}{l}{$C<0$}\\
	\hline
	$Y$     &$Y_0$        & $\cdots$ & $\infty$  \\ \hline 
	$f(aY)$ & $0$ & $\cdots$  & $1$ \\ \hline 
	$\tilde{\Phi}$ & $+\infty$ &   & $-\infty$ \\  
      $\tilde{R}$ &  $\ast$  &   & $0$ \\ \hline
\end{tabular}
\end{center}
\label{dependence}
\end{table}

Here we comment on the case of $a=0$.
Substituting (\ref{calabi}) into (\ref{dual_metric}) and (\ref{dual_dilaton}), 
we obtain the expression of the dual metric and dilaton.
The dual dilaton is no longer a constant.
The curvature is also calculated from (\ref{dual_r}) and  
its dependence on $Y$ can be discussed  as above.

\section{Conclusions}
\label{conclusion}

We derived non-compact K\"ahler backgrounds as a solution of the Einstein equation.
The backgrounds are interpreted as a complex line bundle over a base manifold comprising of
a combination of arbitrary coset spaces.
They have a non-constant dilaton field in general, 
and come to be Calabi-Yau manifolds when a parameter vanishes.
If the base coset space is designated, we can give an explicit formula of the total space metric. 

Furthermore, we obtained a non-K\"ahlerian solution
by performing the duality transformation with respect to the $U(1)$ isometry.
The dual background is equivalent to the original one as a conformal field theory, 
but has a different form as an $N=2$ supersymmetric $\sigma$-model.

\section*{Acknowledgments}

The work of K.S. is supported by the Grant-in-Aid from the Ministry of Education, Science, Sports and 
Culture of Japan ($\sharp$ $14740115$).

\end{document}